\documentclass[conference]{IEEEtran}
\IEEEoverridecommandlockouts
\usepackage{cite}
\usepackage{amsmath,amssymb,amsfonts}
\usepackage{algorithmic}
\usepackage[linesnumbered,ruled,vlined]{algorithm2e}
\usepackage{graphicx}
\usepackage{textcomp}
\usepackage{xcolor}
\usepackage{graphicx}
\usepackage{float}
\usepackage{subfig}
\usepackage{xcolor}
\usepackage{comment}

\makeatletter

\newcommand{\Rmnum}[1]{\expandafter\@slowromancap\romannumeral #1@}
\makeatother
\def\BibTeX{{\rm B\kern-.05em{\sc i\kern-.025em b}\kern-.08em
T\kern-.1667em\lower.7ex\hbox{E}\kern-.125emX}}

\begin{document}

\title{Explainable Deep Learning Based Adversarial Defense for Automatic Modulation Classification}

\author{\IEEEauthorblockN{
	Peihao~Dong,~\IEEEmembership{Member,~IEEE}, Jingchun~Wang, Shen~Gao, Fuhui~Zhou,~\IEEEmembership{Senior Member,~IEEE}, \\
	and Qihui Wu,~\IEEEmembership{Fellow,~IEEE}
}

\vspace{-0.4cm}
\thanks{
	%\vspace{-0.3cm}
	%This work was supported in part by the National Science Foundation	of China under Grants 62231015, 62471226, 62401288, 62101253 and the Natural Science Foundation of Jiangsu Province under Grants BK20240618	and BK20210283. \emph{(Corresponding authors: Peihao Dong.)}
	
	P. Dong, J. Wang, and Q. Wu are with the College of Electronic and Information Engineering, Nanjing University of Aeronautics and Astronautics, Nanjing 211106, China (e-mail: phdong@nuaa.edu.cn; wjc797952@nuaa.edu.cn; wuqihui2014@sina.com).
	
	S. Gao is with the College of Telecommunications and Information Engineering, Nanjing University of Posts and Telecommunications, Nanjing 210003, China (e-mail: gaoshen@njupt.edu.cn).
	
	F. Zhou is with the College of Artificial Intelligence, Nanjing University of Aeronautics and Astronautics, Nanjing 211106, China (e-mail: zhoufuhui@ieee.org).
	
	%Copyright (c) 2025 IEEE. Personal use of this material is permitted. However, permission to use this material for any other purposes must be obtained from the IEEE by sending a request to pubs-permissions@ieee.org.
}
%\vspace{-0.3cm}
}

%\vspace{-0.5cm}
\IEEEtitleabstractindextext{%
\begin{abstract}
	Deep learning (DL) has been widely applied to enhance automatic modulation classification (AMC). However, the elaborate AMC neural networks are susceptible to various adversarial attacks, which are challenging to handle due to the generalization capability and computational cost. In this article, an explainable DL based defense scheme, called SHapley Additive exPlanation enhanced Adversarial Fine-Tuning (SHAP-AFT), is developed in the perspective of disclosing the attacking impact on the AMC network. By introducing the concept of cognitive negative information, the motivation of using SHAP for defense is theoretically analyzed first. The proposed scheme includes three stages, i.e., the attack detection, the information importance evaluation, and the AFT. The first stage indicates the existence of the attack. The second stage evaluates contributions of the received data and removes those data positions using negative Shapley values corresponding to the dominating negative information caused by the attack. Then the AMC network is fine-tuned based on adversarial adaptation samples using the refined received data pattern. Simulation results show the effectiveness of the Shapley value as the key indicator as well as the superior defense performance of the proposed SHAP-AFT scheme in face of different attack types and intensities.
\end{abstract}

}

\maketitle

\IEEEdisplaynontitleabstractindextext

% \fi
\IEEEpeerreviewmaketitle

\section{Introduction}

As wireless devices proliferate, automatic modulation classification (AMC), originally developed for military use, has been widely adopted in civil scenarios such as illegal signal surveillance, internet of things (IoT), emergency communications, and cellular network optimization \cite{b0,b1,b01}, enabling modulation recognition without transmitter–receiver cooperation.

Traditional AMC methods rely on feature extraction, statistical pattern recognition, and decision theory \cite{b2,b3}. To address performance degradation under non-ideal channels and reduce complexity, deep learning (DL) has been employed for automatic feature extraction via nonlinear mapping \cite{b4,b5,b6,b7,b07}. However, DL models are highly vulnerable to adversarial attacks, where minor input perturbations can severely compromise AMC performance \cite{b8}. Attackers may exploit this to disrupt signal recognition by imperceptibly manipulating received signals.

Adversarial attacks are commonly divided into white-box and black-box types based on access to model knowledge. White-box attacks use model parameters and gradients to generate perturbations, including gradient-based (FGSM \cite{b8}, BIM \cite{b9}, PGD \cite{b10}), decision boundary-based (DeepFool \cite{b11}), and optimization-based methods (C\&W \cite{b12}). In contrast, black-box attacks do not require parameter access and include score-based (ZOO, AutoZOO \cite{b13,b14}), query-based (P-RGF), and universal attacks (UAP \cite{b15}).
To mitigate such adversarial effects, many defense strategies have been explored, including input feature-aided methods that enhance robustness.

Explainable artificial intelligence (XAI) offers new directions for adversarial defense by uncovering causal inference paths in DL models through visualization, reasoning, and transparency, thereby addressing DL's black-box nature \cite{b24}. Among various XAI methods, Shapley additive explanation (SHAP) \cite{b25} stand out for its theoretical rigor and practical maturity. SHAP quantifies each input’s contribution to model output via Shapley values, a positive Shapley value indicates that the feature contributes positively to the network prediction while a negative Shapley value indicates a destructive contribution, and magnitudes indicate importance. Therefore, Shapley values can reflect the destructive impact of adversarial attacks on the task, thus guiding improvements in model performance.

\subsection{Related Works}

In the field of AMC, an increasing number of sophisticated adversarial methods building upon traditional white-box attacks have been developed in recent years.
A dual iterative loop method was proposed in \cite{b16}, which dynamically updates initial conditions in each iteration to enhance the attack adaptability. In \cite{b17}, a broadcast adversarial attack method incorporating channel effects was developed to generate universal perturbations for different receivers, thereby increasing the overall effectiveness of the attack.

To cope with the increasingly sophisticated adversarial attacks, a growing number of advanced defense strategies have been proposed. In \cite{b18}, an information bottleneck mechanism was used to extract task-relevant features. Data mapping and noise filtering with signal enhancement were respectively adopted in \cite{b19} and \cite{b20} to improve the model’s defense capability. Knowledge distillation was leveraged to mitigate the adversarial impact via a multi-level distillation mechanism \cite{b21} and a defensive distillation approach \cite{b22}. A multi-network incorporated ensemble learning strategy was elaborated in \cite{b23} to reduce the impact of adversarial perturbations. Furthermore, several extended approaches were also developed to enhance the adversarial robustness of AMC networks, including the attack detection aided misclassification correction \cite{b231}, adversarial sample purification \cite{b232}, and frequency-domain homomorphic filtering based perturbation suppression \cite{b233}.

The emergence of XAI opens up new avenues for the adversarial defense, which has been widely applied to the medical image analysis, credit risk assessment, and autonomous vehicle decision-making due to its unique ability to address the black-box crux plaguing DL. For DL-based AMC, it is also essential to interpret the model always treated as a black box to improve the recognition performance. An interpreting framework, named RobustRMC, was proposed in \cite{b241} to explain the convolutional neural network (CNN) and long short-term memory (LSTM) architectures via the visualization method, demonstrating that the adversarial training helps models refocus on key feature regions and thus improves the overall robustness. In \cite{b242}, an explainable approach combining the traditional feature extraction with a decision tree surrogate model was proposed, which enhances the stability and local fidelity of the generated explanations by incorporating the deterministic neighborhood sampling and a stable oversampling mechanism. Aiming to make explanations align with the human intuition, a model-agnostic interpretable method was proposed in \cite{b243}, which assigns importance scores to each time point in the signal.

Several variants of SHAP were applied to develop attack detection methods \cite{b26,b27}, demonstrating that SHAP-based classifiers not only enhance the model interpretability but also improve the detection performance for the cybersecurity threat. In \cite{b28}, an SHAP-based supervised clustering method was proposed for the explainable fault analysis in mobile networks, which facilitates the identification of fault clusters and the formulation of corresponding mitigation strategies. In \cite{b29}, a pattern-dependent SHAP was developed to interpret the LSTM predictions by revealing the underlying traffic patterns for both correct and incorrect classifications. In \cite{b30}, four mainstream DNN explanation methods including SHAP were applied to evaluate the contribution of input signals with respect to the predictions of an AMC network.

\subsection{Motivation and Contributions}

Despite significant advances, the defense method for DL-based AMC still needs to be improved in terms of the model generalization, robustness against different types of attacks, and training complexity. Specifically, a fine-grained explanation of the attack's impact mechanism is necessary through providing the quantitative analysis and targeted correction for those destructive features introduced by attacks. In addition, enhancing the generalization capability of defense methods is critical to cope with various types of adversarial attacks. The problem of complicated and intractable parameter optimization should also be well treated. To effectively mitigate the impact of adversarial attacks on DL-based AMC, a novel SHAP-based defense framework is developed in this article. The main novelty and contribution can be summarized
as follows:
\begin{itemize}
\item Shapley value is first exploited to disclose the impact of adversarial attacks on the AMC network quantitatively, based on which SHAP is considered as the potential defense solution. The cognitive negative information is formulated to depict the widespread adverse effect in the information cognition. By resorting to the properties of Shapley value, the motivation of using SHAP for defense is theoretically analyzed.

\item A general defense scheme, SHAP-enhanced adversarial fine-tuning (SHAP-AFT), is proposed to cope with various types of attacks, which is composed of the attack detection, SHAP-based information importance evaluation, and AFT stages. By designing a compact AMC network and using only a few online collected samples, the computational cost of building the SHAP explainer is reduced. Furthermore, the AFT is conducted based on the offline generated samples to lighten the load of collecting samples online.

\item Extensive simulation experiments are conducted to verify the effectiveness of the proposed SHAP-AFT defense scheme. The consistence between the Shapley heatmap and the classification confusion matrix justifies SHAP for the adversarial defense. Compared with baseline schemes, SHAP-AFT exhibits the superior classification accuracy and robustness, especially under the high intensity of attacks.
\end{itemize}
%The first stage indicates the existence of the attack. The second stage evaluates contributions of the received data and removes those data positions with negative Shapley values corresponding to the dominating negative information caused by the attack. Then the AMC network is fine-tuned based on adversarial adaptation samples using the refined received data pattern.

\subsection{Organization}

The rest of the paper is organized as follows. Section \Rmnum{2} describes the signal transmission and adversarial attack models. Section \Rmnum{3} analyzes the theoretical motivation and feasibility of using SHAP to defend attacks, based on which Section \Rmnum{4} elaborates the proposed SHAP-AFT framework. Section \Rmnum{5} demonstrates the effectiveness of the SHAP-AFT framework via extensive simulation results. Finally, Section \Rmnum{5} concludes this article.

\section{Signal Transmission and Attack Model}

\subsection{Signal Transmission Model}
In the AMC system, the receiver recognizes the modulation type of the received signal for purpose of surveillance or demodulation. The modulated signal at the transmitter is represented as $\mathbf{s}=[s[1],...,s[L]]$, where $L$ is the signal length.
After going through the wireless channel, the signal received at the $l$th time instant can be expressed as
\begin{equation}
x[l]=h[l]e^{-j2\pi(vlT_s+\theta)}s[l]+z[l],  l=1,...,L \label{eq}
\end{equation}
where $h[l]$ and $z[l]$ respectively denote the channel gain and additive white Gaussian noise (AWGN) at the $l$th time instant with $T_s$, $v$, and $\theta$ representing the sampling period, frequency shift, and phase shift, respectively. The receiver needs to determine the modulation type of the signal, $C(s)$, from the candidate set, $\mathcal{M}$, including $M$ elements. Denote $\mathbf{x}=[x[1],...,x[L]]$ as the vector form of the received signal. Then the input for the AMC network can be expressed as
\begin{equation}
\mathbf{X} = \left[ \Re\{\mathbf{x}\},\Im\{\mathbf{x}\}\right] ,
\end{equation}
where $\mathbf{x}=[x[1],...,x[L]]^T$, $\Re\{\cdot\}$ and $\Im\{\cdot\}$ denote the real and imaginary parts, respectively.

\subsection{Attack Models}
\subsubsection{Fast Gradient Sign Method}
DL-based AMC is vulnerable to various types of network attacks, which can significantly degrade its classification accuracy. Let the original input be denoted by $\mathbf{x}$, and the adversarial input, generated by adding a perturbation $\mathbf{\eta}$, be represented as $\mathbf{\tilde{x}}$. In DL-based multi-classification tasks, the training process aims to minimize the loss function $\mathcal{L}$. Adversarial attacks amplify the loss function by introducing a slight perturbation $\mathbf{\eta}$ to the input, resulting in the classification errors
\begin{equation}
	\mathop{\arg\max}\limits_{\theta}\mathcal{L}(f(\theta,\mathbf{x}+\mathbf{\eta}),\mathbf{y}), \label{eq}
\end{equation}
where \(f(\cdot)\) denotes the network, $y$ is the label for input $\mathbf{x}$, and $\theta$ represents parameters requiring optimization.
The fast gradient sign method (FGSM) optimizes perturbation $\mathbf{\eta}$ through gradient ascent during back propagation. The formula is
\begin{equation}
	\eta=\varepsilon sign(\nabla_{x}\mathcal{L}(\theta,\mathbf{x},\mathbf{y})), \label{eq}
\end{equation}
where $\nabla_{x}\mathcal{L}$ is the partial derivative of the loss function $\mathcal{L}$ with respect to input $\mathbf{x}$, $sign(\cdot)$ denotes the sign function, and $\varepsilon$ is the perturbation step size. The adversarial samples $\mathbf{\tilde{x}}$ generated by FGSM are expressed as
\begin{equation}
	\mathbf{\tilde{x}}=\mathbf{x}+\mathbf{\eta}. \label{eq}
\end{equation}

\subsubsection{Basic Iterative Method}
The basic iterative method (BIM) extends the FGSM by applying perturbations iteratively rather than in a single step. In the $n$-th iteration of the attack, $\mathbf{\tilde{x}}_{n+1}$, is generated by applying the perturbation to the previous adversarial sample $\mathbf{\tilde{x}}_{n}$. To ensure that the generated adversarial samples remain within the predefined perturbation range $\varepsilon$, the updated sample $\mathbf{\tilde{x}}_{n+1}$ is clipped after each iteration to the interval $[\mathbf{x} - \varepsilon, \mathbf{x} + \varepsilon]$. The final adversarial sample is thus given by
\begin{equation}
	\mathbf{\tilde{x}}_{n+1} = \operatorname{Clip}_{x, \varepsilon} \left\{ \mathbf{\tilde{x}}_n + \alpha \cdot \operatorname{sign} (\nabla_x \mathcal{L}(\theta, \mathbf{\tilde{x}}_n, \mathbf{y})) \right\}.
\end{equation}

\begin{figure}[!t]
	\centering
	\includegraphics[width=0.45\textwidth]{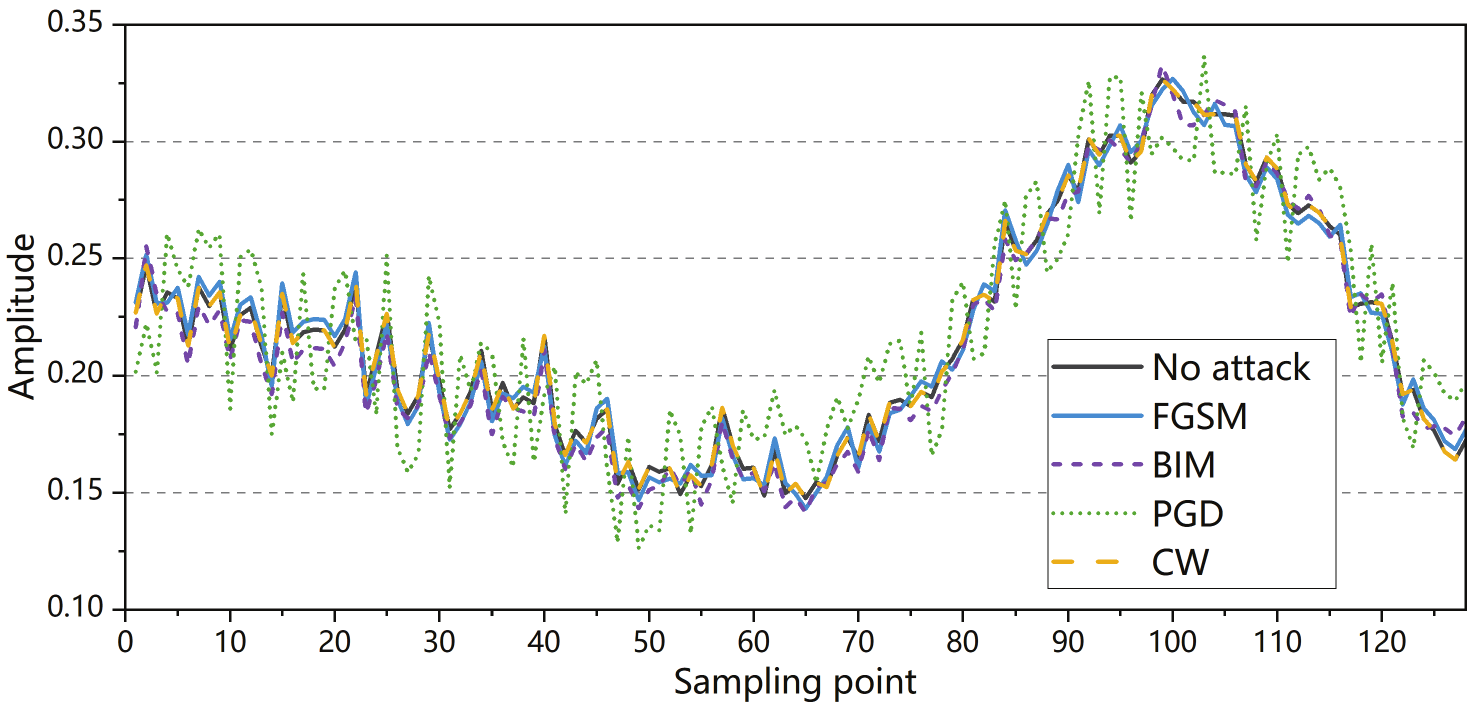}
	\caption{Amplitude of WBFM modulation under adversarial attacks.}
	\label{figattack}
\end{figure}

\subsubsection{Projected Gradient Descent Method}
In nonlinear networks, the perturbation direction may vary across iterations. To address this, the projected gradient descent (PGD) attack employs iterative projection to enhance attack effectiveness. The PGD adversarial sample $\mathbf{\tilde{x}}_{n+1}$ is given by
\begin{equation}
	\mathbf{\tilde{x}}_{n+1} = Proj_{x,\varepsilon}\{\mathbf{\tilde{x}}_n+\alpha \cdot \operatorname{sign} (\nabla_x \mathcal{L}(\theta, \mathbf{\tilde{x}}_n, \mathbf{y}))\}.
\end{equation}
Here, $Proj_{x,\varepsilon}$ denotes the projection operation, which ensures that the perturbation $\mathbf{\eta_n}$ satisfies $\lVert \eta \rVert_{\infty} < \varepsilon$.

\subsubsection{Carlini \& Wagner Attack}

The Carlini \& Wagner (C\&W) attack generates adversarial samples through an optimization process. The objective is to minimize the difference between the adversarial and original samples while ensuring that the adversarial sample is misclassified. This optimization problem is formulated as
\begin{equation}
	\min D(\mathbf{x},\mathbf{x}+\eta),
\end{equation}
subject to
\begin{equation}
	C(\mathbf{x} + \mathbf{\eta}) = t, \mathbf{\quad x} + \mathbf{\eta} \in [0,1]^n,
\end{equation}
here, $D$ denotes a distance function that quantifies the difference between  $\mathbf{x}$ and $\mathbf{\tilde{x}}$, $C$ represents the classifier, and $t$ is the target class to which the adversarial sample is intended to be misclassified.

Fig. \ref{figattack} illustrates the amplitude variations of the WBFM-modulated signal under different types of adversarial attacks, which is misclassified as the 64QAM modulation. It is noted that FGSM, BIM and C\&W attacks incur imperceptible impacts with the perturbative signal amplitudes almost same as the original signal so that the defense difficulty is aggravated.
%Among the four attack types, the C\&W attack incurs the most imperceptible impact with the perturbative signal amplitude almost same as the original signal.

\section{Theoretical Motivation of Using SHAP for Defense}

In this section, the cognitive negative information is formulated to depict the widespread adverse effect in the information cognition. Then the unique ability of SHAP in finding out the negative information and thus defending the adversarial attack is analyzed.

\subsection{Cognitive Negative Information}

According to the traditional information theory, the amount of information of Event $X$ can be only increased or maintained conditioned on Event $Y$, which is formulated by the well-known mutual information as
\begin{eqnarray}
\label{MI}
I(X;Y)=H(X)-H(X| Y),
\end{eqnarray}
where $H(X)$ and $H(X| Y)$ measure the uncertainties of Event $X$ itself and conditioned on Event $Y$, respectively. The mutual information satisfies $I(X;Y) \ge 0$ as $H(X)\ge H(X| Y)$ always holds true. This fundamental formula directs the information transmission. However, in the information cognition, an information processing mode more approximated to the human brain, the occurrence of Event $Y$ may increase the uncertainty of Event $X$. In other words, it is ubiquitous in cognitive scenarios that Event $Y$ representing the interference, attack, and deception will increase the cognitive difficulty of Event $X$ representing the original cognitive task, which, however, can not be formulated by the mutual information in (\ref{MI}). Therefore, it is necessary to construct the metric formula matched with the information cognition.

Denote the cognitive difficulties of Event $X$ itself and conditioned on Event $Y$ as $D(X)$ and $D(X| Y)$, respectively. The cognitive information can be expressed as
\begin{eqnarray}
\label{CI}
C(X;Y)=D(X)-D(X| Y),
\end{eqnarray}
where both $D(X)\ge D(X| Y)$ and $D(X)< D(X| Y)$ can hold true. Specifically, we have $C(X;Y)< 0$ as $Y$ increases the cognitive difficulty of $X$, i.e., $D(X)< D(X| Y)$, in which case the \emph{cognitive negative information} is introduced by Event $Y$.

In this article, $X$ and $Y$ respectively represent the original AMC task and the adversarial attack, where $Y$ introduces the negative information and increases the cognitive difficulty of $X$. In the following, SHAP will be exploited to reduce the negative information and thus to defend attacks by analyzing the cognitive difficulty.

\subsection{Reducing Negative Information and Defending via SHAP}

For the DL-based AMC, the classification network maps the in-phase/quadrature (I/Q) data to the probabilities of candidate modulation types and selects that with the maximum probability. The cognitive difficulty of $X$, $D(X)$, is positively correlated to the loss of the AMC network, which is expressed as
\setlength{\arraycolsep}{0.1em}
\begin{eqnarray}
\label{CD_loss}
D(X)\propto \mathcal{L}&&=-\sum_{n=1}^{N}\sum_{k=1}^{M} y^{n,k} \log(\hat{y}^{n,k})\nonumber\\
&&\stackrel{(a)}=-\sum_{n=1}^{N} \log(\hat{y}^{n,k_{0}}),
\end{eqnarray}
where $y^{n,k}\in \{0,1\}$ and $\hat{y}^{n,k}\in [0,1]$ denote the label value and prediction value of the $k$th modulation type for the $n$th sample, respectively, while $k_{0}$ denotes the index of the true modulation type. Then we have $y^{n,k}=1$ for $k=k_{0}$ and $y^{n,k}=0$ otherwise, yielding a compact form via $(a)$. From (\ref{CD_loss}), the cognitive difficulty, $D(X)$, will increase as $\hat{y}^{n,k_{0}}$ decreases.

It is difficult to analyze the model prediction value, $\hat{y}^{n,k_{0}}$, directly. According to the property of additive feature attribution methods, $\hat{y}^{n,k_{0}}$ can be approximately decomposed by an explanation model via a linear form as
\begin{eqnarray}
\hat{y}^{n,k_{0}}\approx g(\mathbf{z}) = \Phi_0 + \sum_{i=1}^{L} \Phi_i z_i',
\end{eqnarray}
where $\mathbf{z}=[z_1',\dots,z_L']$ denotes the simplified binary vector indicating whether the input features are included, $\Phi_0$ and $\Phi_i$ represent the baseline output without any input features and the contribution of the $i$th input feature, respectively.

\begin{figure*}[htbp]
\centering
\includegraphics[width=0.92\textwidth]{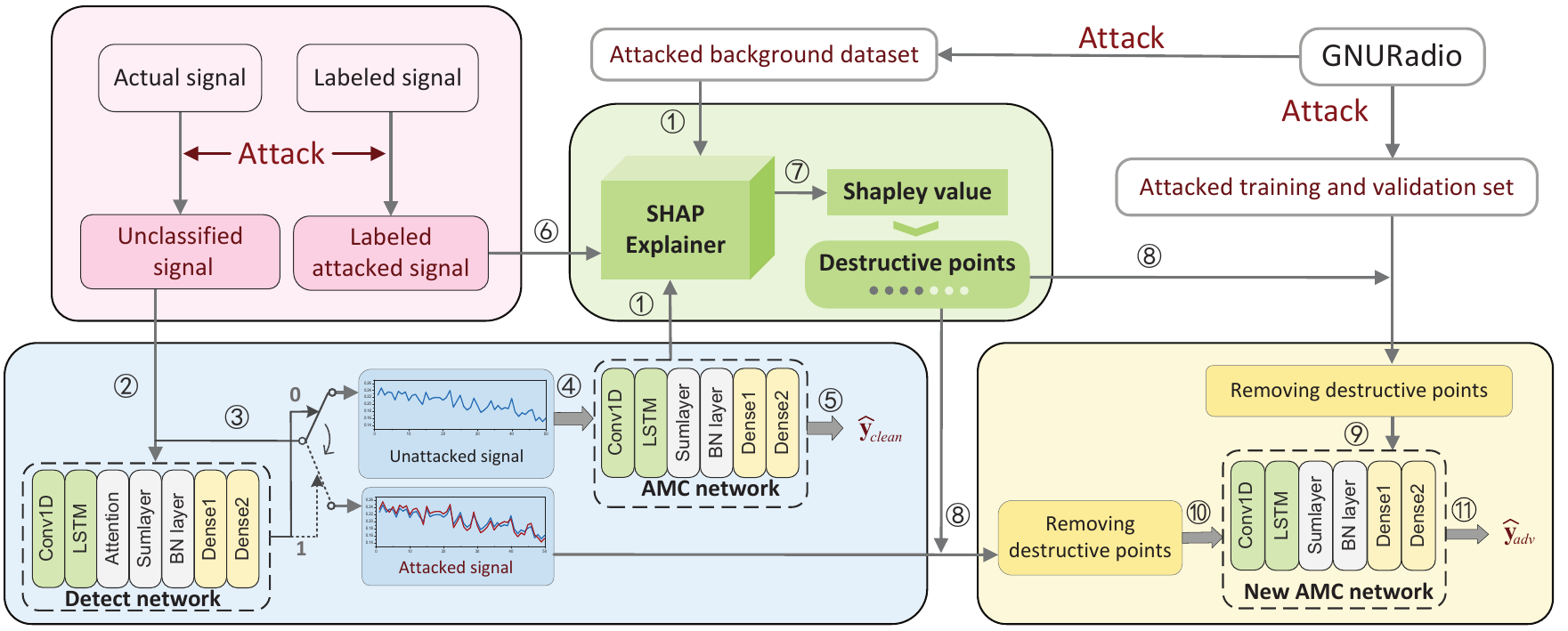}
\caption{SHAP-AFT algorithm framework.}
\label{SHAP-AFT}
\end{figure*}

The Shapley value, originating from game theory, is one of the most general metrics to measure the feature contribution. Specifically, $\Phi_i$ can be decomposed into the real-part Shapley value $\Phi_{i^{\Re}}$ corresponding to $\Re\{x[i]\}$ and imaginary-part Shapley value $\Phi_{i^{\Im}}$ corresponding to $\Im\{x[i]\}$, that is
\begin{eqnarray}
\Phi_i =\Phi_{i^{\Re}}+\Phi_{i^{\Im}},
\end{eqnarray}
where $i^{\Re}$ and $i^{\Im}$ denote the indices of the real and imaginary parts of the $i$th received data point, respectively. Then we have
\setlength{\arraycolsep}{0.1em}
\begin{eqnarray}
\Phi_{i^{\Re}} = \sum_{\mathcal{S} \subseteq \mathcal{F} \setminus \{i^{\Re}\}} \frac{|\mathcal{S}|!(|\mathcal{F}|\! -\! |\mathcal{S}| \!-\! 1)!}{|\mathcal{F}|!} \left[ f(\mathbf{x}_{\mathcal{S} \cup \{i^{\Re}\}}) - f(\mathbf{x}_{\mathcal{S}}) \right],\nonumber\\
\Phi_{i^{\Im}} = \sum_{\mathcal{S} \subseteq \mathcal{F} \setminus \{i^{\Im}\}} \frac{|\mathcal{S}|!(|\mathcal{F}|\! -\! |\mathcal{S}| \!-\! 1)!}{|\mathcal{F}|!} \left[ f(\mathbf{x}_{\mathcal{S} \cup \{i^{\Im}\}}) - f(\mathbf{x}_{\mathcal{S}}) \right],\nonumber\\
\end{eqnarray}
where $\mathcal{F}$ and $\mathcal{S}$ represent the set including all features and an arbitrary subset, respectively. Furthermore, $f(\mathbf{x}_{\mathcal{S} \cup \{i^{\Re/\Im}\}})$ and $f(\mathbf{x}_{\mathcal{S}})$ denote the output of the original model with input features $\mathcal{S}$ and $i^{\Re/\Im}$ and with only $\mathcal{S}$, respectively, which can be computed by training two models with the matched number of input neurons. However, training models for all possible $\mathcal{S}$ requires the prohibitively high computational cost and is impractical to implement, especially when the feature dimension becomes large. To address this problem, the integrated gradient (IG) is used to approximate the Shapley values. Take the real-part as example, the formula of IG is given by
\setlength{\arraycolsep}{0.1em}
\begin{eqnarray}
\mathrm{IG}_{i^{\Re}}(\mathbf{X}) = (x_{i^{\Re}} - x_{i^{\Re}}') \int_0^1 \frac{\partial f(\mathbf{X}' + \alpha \times (\mathbf{X} - \mathbf{X}'))}{\partial x_{i^{\Re}}} d\alpha,\nonumber\\
\end{eqnarray}
where $\mathbf{X}$ and $\mathbf{X}'$ denote the original and reference input matrices, respectively, with $x_{i^{\Re}}$ and $x_{i^{\Re}}'$ corresponding to the first entry of the $i$th row therein. The scalar $\alpha$ ranging from $0$ to $1$ accounts for the smooth change of the input from  $\mathbf{X}'$ to $\mathbf{X}$. However, it is difficult to find out a universal reference vector $\mathbf{X}'$ for different cases and averaging over multiple reference samples will involve multiple computation-intensive integrals. By introducing the expectation of the IG and then simplifying the integral operation via sampling, the the expected gradient (EG) can be expressed as
\setlength{\arraycolsep}{0.1em}
\begin{eqnarray}
\label{EG}
\mathrm{EG}_{i^{\Re}}(\mathbf{X}) \!=\!\! \underset{\mathbf{X}'\sim D, \atop \alpha\sim \mathcal{U}(0,1)}{\mathbb{E}} \left[(x_{i^{\Re}} - x_{i^{\Re}}') \frac{\partial f(\mathbf{X}' + \alpha(\mathbf{X} - \mathbf{X}'))}{\partial x_{i^{\Re}}} \right],\nonumber\\
\end{eqnarray}
where $\mathcal{U}(0,1)$ denotes the uniform distribution from $0$ to $1$. Then the real-part Shapley value is approximated as $\Phi_{i^{\Re}}\approx \mathrm{EG}_{i^{\Re}}(\mathbf{X})$. The imaginary-part Shapley value can be obtained similarly as $\Phi_{i^{\Im}}\approx \mathrm{EG}_{i^{\Im}}(\mathbf{X})$. With all features input into the model, $\hat{y}^{n,k_{0}}$ can be written as
\begin{eqnarray}
\hat{y}^{n,k_{0}}\approx \Phi_0 + \sum_{i=1}^{L} \Phi_i,
\end{eqnarray}
where $\Phi_i>0, \forall i=1,\ldots,L$ generally holds true in the attack-free case since all received data, $x_1,\ldots,x_L$, have positive effects on the classification result. In other words, all received data are isomorphic and the superposition of their effects is constructive. When the attack is imposed, the prediction value becomes
\begin{eqnarray}
\tilde{y}^{n,k_{0}}\approx \Phi_0 + \sum_{i=1}^{L} \tilde{\Phi}_i=\Phi_0 + \sum_{j\in \mathcal{P}} \tilde{\Phi}_j-\sum_{l\in \mathcal{N}} |\tilde{\Phi}_l|,
\end{eqnarray}
where $\mathcal{P}$ and $\mathcal{N}$ denote the index sets corresponding to the received data with positive and negative contributions, respectively. Some $\tilde{\Phi}_i$s become negative due to the destructive impact caused by the attack. The values of $\tilde{\Phi}_1,\ldots,\tilde{\Phi}_L$ reveal how the received data under attack, $\tilde{x}_1,\ldots,\tilde{x}_L$, influence the classification result and $\tilde{\Phi}_l, l\in \mathcal{N}$, point to those received data with the dominated destructive impacts therein. Then the received data in $\mathcal{P}$ and $\mathcal{N}$ can be regarded as heterogeneous and the superposition of their effects is destructive. Generally, $\hat{y}^{n,k_{0}}>\tilde{y}^{n,k_{0}}$ holds true and corresponds to $D(X)< D(X| Y)$, indicating that $D(X| Y)$ can be decreased by increasing $\tilde{y}^{n,k_{0}}$. Based on the above analysis, we can remove the data with the destructive contribution, i.e.,  $\tilde{\Phi}_l$, from the received data sequence to improve the cognitive accuracy. It is noted that $\mathcal{N}$ usually changes along with the sample index, $n$, and a relatively universal remove criterion for different samples is needed. A simple way is to average $\tilde{\mathbf{\Phi}}=[\tilde{\Phi}_1, \ldots, \tilde{\Phi}_L]$ over all samples, which yields the averaged contribution vector, $\bar{\mathbf{\Phi}}$, with the corresponding destructive contribution set, $\bar{\mathcal{N}}$. Therefore, the theoretical motivation of using SHAP for the adversarial defense is justified.

\section{SHAP-AFT Based Defense Scheme}

In this section, the SHAP-AFT based defense scheme is proposed. The algorithm framework is first reviewed, after which structures of the attack detection network and AMC network are elaborated.

\subsection{Algorithm Framework}

In general, the SHAP-AFT based defense scheme includes three key parts: 1) The attack detection network distinguishes whether the signal is attacked. 2) The destructive data points of the attacked signal are found out according to the Shapley values calculated by the SHAP explainer. 3) The AMC network is fine-tuned using adversarial samples with destructive data points in the input removed.

The algorithm framework is detailed in Fig.~\ref{SHAP-AFT}. Specifically, the attacked received signals are split into three subsets: attacked background dataset for building SHAP explainer, attacked training and validation set for adversarial fine-tuning. First, a small number of labeled signals are transmitted and subjected to adversarial attacks in the the channel propagation, after which the actual signals are transmitted. At the receiver end, two types of attacked signals are obtained: labeled attacked signal samples and unclassified actual signal samples. First, the previously generated attacked background dataset is used as reference data of the AMC network to construct the SHAP explainer (Step\textcircled{\scriptsize{1}}), which employs the EG algorithm in (\ref{EG}) to compute Shapley values. Then the unclassified signal is first fed into the attack detection network (Step\textcircled{\scriptsize{2}}), which determines whether each sample has been subjected to an attack (Step\textcircled{\scriptsize{3}}). If a sample is categorized as the unattacked signal, it is directly forwarded to the AMC network (Step\textcircled{\scriptsize{4}}), resulting in the the corresponding modulated classification label $\hat{\mathbf{y}}_{clean}$ (Step\textcircled{\scriptsize{5}})and an classification accuracy denoted as $A_{clean}$. Next the labeled attacked signal samples are then input into the explainer to obtain their corresponding Shapley values (Step\textcircled{\scriptsize{6}}). Through filtering and sorting, the feature points those carry the negative information and destructively impact the classification performance are identified (Step\textcircled{\scriptsize{7}}). Since both the labeled and unlabeled attacked signals have undergone the same adversarial attack, the destructive features identified from the labeled signal can be used to guide the removal of destructive features in the actual attacked samples (Step\textcircled{\scriptsize{8}}).

However, removing destructive feature points from the attacked signal samples changes the signal dimension, making it incompatible with the original input format of the AMC network. To address this, the refined attacked training and validation set with the changed input dimension are used to fine-tune the AMC network through adversarial training (Step\textcircled{\scriptsize{9}}), resulting in a newly adapted and more robust AMC network.
The newly trained network is not only adapted to the structure of the refined attacked signals but also demonstrates enhanced defensive capability as a result of adversarial training. Finally, the refined attacked signals, i.e., attacked signals with destructive data points removed, are input into the new AMC network (Step\textcircled{\scriptsize{10}}) for modulation classification, yielding the the corresponding modulated classification label $\hat{\mathbf{y}}_{adv}$ (Step\textcircled{\scriptsize{11}}) and an accuracy denoted as $A_{adv}$. The overall performance of SHAP-AFT is then calculated as a weighted sum of $A_{clean}$ and $A_{adv}$, that is,
\begin{eqnarray}
A_{oa}=\beta_{clean}A_{clean}+\beta_{adv}A_{adv},
\end{eqnarray}
where $\beta_{clean}$ and $\beta_{adv}$ represent the proportions of clean and attacked signal samples, respectively, after the unclassified samples are processed by the attack detection network. The SHAP-AFT based defense scheme is summarized in Algorithm 1.

\begin{algorithm}[htbp]
\caption{The SHAP-AFT Algorithm}
\KwIn{Unclassified signal $\mathbf{x}$, labeled signal $\mathbf{x}_{lab}$, background data $\mathbf{x}_{bg}$, the training set $\mathbf{x}_{tr}$, the validation set $\mathbf{x}_{val}$ for fine-tuning, and the index set of destructive data points $\mathcal{M}_{des} = \emptyset$.}
\KwOut{Modulation classification results $\hat{\mathbf{y}}_{clean}$ of the AMC network or $\hat{\mathbf{y}}_{adv}$ of the new AMC network.}
Generate the attacked data $\tilde{\mathbf{x}}, \tilde{\mathbf{x}}_{lab}, \tilde{\mathbf{x}}_{train}, \tilde{\mathbf{x}}_{val}, \tilde{\mathbf{x}}_{bg}$\;
Construct SHAP Explainer: $SE(f_{AMC},\tilde{\mathbf{x}}_{bg})$\;
\ForEach{$\tilde{\mathbf{x}} \in \tilde{\mathcal{X}}$}{
	Compute $f_{detect}(\tilde{\mathbf{x}}) \in \{0,1\}$\;
	\eIf{$f_{detect}(\tilde{\mathbf{x}}) = 0$}{
		$\tilde{\mathbf{x}} = \tilde{\mathbf{x}}_{clean}$\;
		$\hat{\mathbf{y}}_{clean} = f_{AMC}(\tilde{\mathbf{x}}_{clean})$\;
	}{
		$\tilde{\mathbf{x}} = \tilde{\mathbf{x}}_{adv}$\;			
		\ForEach{$\tilde{\mathbf{x}}_{lab} \in \tilde{\mathcal{X}}_{lab}$}{
			$\tilde{\mathbf{\Phi}}=SE(f_{AMC},\tilde{\mathbf{x}}_{lab})=[\tilde{\Phi}_1, \ldots, \tilde{\Phi}_L]$\;
			%Assign to $\tilde{\boldsymbol{\Phi}}$.
		}
		
		Average $\tilde{\mathbf{\Phi}}=[\tilde{\Phi}_1, \ldots, \tilde{\Phi}_L]$ over all samples to obtain the averaged contribution vector, $\bar{\mathbf{\Phi}}$\;
		Sort $\bar{\boldsymbol{\Phi}}$ in an ascent order and obtain the sub-vector $\bar{\boldsymbol{\Phi}}_{neg}$ with negative Shapley values\;
		Select $m$ elements with the smallest values from $\bar{\boldsymbol{\Phi}}_{neg}$ and put their indices into $\mathcal{M}_{des}$\;
		Refine $\tilde{\mathbf{x}}_{train}, \tilde{\mathbf{x}}_{val}, \tilde{\mathbf{x}}_{adv}$ by removing the data points according to the indices in $\mathcal{M}_{des}$, yielding $\tilde{\mathbf{x}}^*_{train}, \tilde{\mathbf{x}}^*_{val}, \tilde{\mathbf{x}}^*_{adv}$\;
		Use $\tilde{\mathbf{x}}^*_{train}, \tilde{\mathbf{x}}^*_{val}$ to conduct the adversarial fine-tuning and obtain new AMC network $f^*_{AMC}$\;
		$\hat{\mathbf{y}}_{adv} = f^*_{AMC}(\tilde{\mathbf{x}}^*_{adv})$\;
		
	}
}	
\Return $\hat{\mathbf{y}}_{clean}$ or $\hat{\mathbf{y}}_{adv}$\;
\end{algorithm}

\subsection{Network Structures}
\subsubsection{Attack Detection Network}
The function of the attack detection network is to determine whether an input signal sample has been subjected to an adversarial attack. If the sample is classified as clean, it is directly forwarded to the AMC network whose structure will be detailed for modulation classification. Otherwise, the SHAP method is applied to analyze and filter out the destructive data points prior to classification.

To sufficiently capture the underlying correlation in the I/Q signal sequence, the proposed network organically conflates the one-dimensional convolutional (Conv1D), long short-term memory, self-attention, and fully connected layers.
At the beginning, a Conv1D layer with 128 kernels of length 8 and the rectified linear unit (ReLU) activation function is first applied to process the input $\mathbf{X} \in \mathbb{R}^{L \times 2}$, generating the feature map $\mathbf{H}_{1} \in \mathbb{R}^{(L-7) \times 128}$.
Then, an LSTM layer is adopted by considering both the local and global dependencies, which includes $L-7$ LSTM units and the output dimension of each unit is $128$. By collecting the output of all units, the feature map $\mathbf{H}_{2} \in \mathbb{R}^{(L-7) \times 128}$ is obtained.
Considering the inherent problem of forgetting of LSTM, a self-attention layer is then appended to enable the network to learn dependencies among different time steps within the sequence, which enhances the contextual representation of the feature map. To reduce the computational complexity, the query (Q), key (K), and value (V) matrices are given by
\begin{equation}
\mathbf{Q}=\mathbf{K}=\mathbf{V}=\mathbf{H}_{2}.
\end{equation}
Then, the importance matrix is calculated via the scaled dot-product as
\begin{equation}
\mathbf{M}=\mathsf{Softmax}\left( \frac{\mathbf{Q} \mathbf{K}^T}{\sqrt{d}}\right),
\end{equation}
based on which the feature matrix, $\mathbf{F} \in \mathbb{R}^{(L-7) \times 128}$, is given by
\begin{equation}
\mathbf{F}=\mathbf{M} \mathbf{V}.
\end{equation}
To obtain a global representation, a sum layer is applied by summing over all time steps as
\begin{equation}
\bar{\mathbf{f}} = \sum_{t=1}^{L-7} \mathbf{f}_t \in \mathbb{R}^{d},
\end{equation}
where $\mathbf{f}_t$ denotes the $t$th row of $\mathbf{F}$. The feature vector $\mathbf{\bar{f}}$ is first processed by a batch normalization (BN) layer before going through two dense layers. The first dense layer contains 256 units with ReLU activation function, and the second dense layer contains 1 unit with softmax activation function to output the final output $\hat{p}\in [0,1]$. Then the signal is categorized as attacked if $\hat{p}\geq 0.5$ holds true and as unattacked otherwise. The loss function can be written as %BCE
\begin{equation}
\mathcal{L}_{\text{BCE}} = -\frac{1}{N} \sum_{i=1}^{N} \left[ p_i \log(\hat{p}_i) + (1 - p_i)  \log(1 - \hat{p}_i) \right],
\end{equation}
where $N$ denotes the number of training samples, $p_i$ and $\hat{p}_i$ denote the label and prediction of the $i$th sample, respectively.

\subsubsection{AMC Network}
Similar to the attack detection network, the convolutional and LSTM layers are jointly used to extract the spatiotemporal features from the input signal $\mathbf{X} \in \mathbb{R}^{L \times 2}$. The output matrix $\mathbf{H} \in \mathbb{R}^{(L-7) \times 128}$ from the LSTM layer is then passed to a sum layer to realize the dimensionality reduction, that is
\begin{equation}
\bar{\mathbf{h}} = \sum_{t=1}^{L-7} \mathbf{h}_t \in \mathbb{R}^{128},
\end{equation}
where $\mathbf{h}_t$ denotes the $t$th row of $\mathbf{H}$.
The resulting feature vector $\bar{\mathbf{h}}$ is subsequently fed into a BN layer to stabilize the training process and accelerate convergence. After that, a dense layer with 256 neurons and ReLU is applied to perform nonlinear feature transformation. Then the output dense layer maps the feature vector to a probability vector $\hat{\mathbf{y}}\in\mathbb{R}^{M}$ indicating the approximate likelihood of each candidate modulation type. The loss function of the AMC network is defined as (\ref{CD_loss}).

\section{Simulation Results}
\subsection{Simulation Setups}\label{AA}
The experiment uses the RML2016.10a signal modulation dataset, which is generated in a GNU Radio + Python environment. This dataset simulates real-world channel effects, including additive white Gaussian noise, channel fading, carrier frequency offset, and sampling rate offset.

The dataset includes eight digital and three analog modulation types: 8PSK, BPSK, CPFSK, GFSK, PAM4, 16QAM, 64QAM, QPSK, AM-DSB, AM-SSB, and WBFM. The signal-to-noise ratio (SNR) ranges from -20 dB to 18 dB with 2 dB step size. The data sample is in IQ format, with the shape of $128\times2$.

The dataset contains 220,000 samples, and is divided into training, validation, and test sets in a 3:1:1 ratio for AMC network training. The dataset used for training the attack detection network consists of 35,200 samples, and is also divided into training, validation, and test sets in a 3:1:1 ratio.
330 samples are selected as labeled signal, and their labels are recorded for later SHAP analysis to identify the data points including the negative information. A small subset of samples is selected by sorting the Shapley values of the feature data points under adversarial attacks, from which 330 samples are used as attacked background dataset for constructing the gradient explainer, 2,000 samples are used as attacked training set, and 400 samples are used as attacked validation set for online adversarial training. Additionally, the experiments related to labeled signals and actual signals are conducted using signal samples with SNR greater than 7 dB.

%\subsection{Shapley Value Processing}
\begin{figure}[htbp]
\centering
\includegraphics[width=0.45\textwidth]{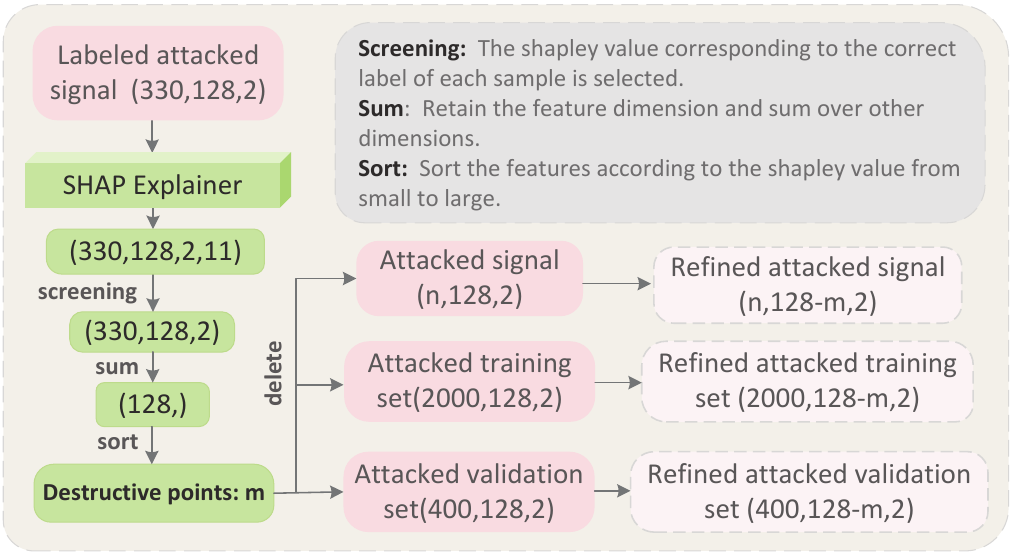}
\caption{Shapley value processing.}
\label{figdatapro}
\end{figure}
For the SHAP computation and analysis process in Fig. \ref{figdatapro}, the AMC network and background data construct a gradient explainer, which allocates feature contributions based on Shapley values. The detailed data processing steps are illustrated in Fig. \ref{figdatapro}.
When the input labeled attacked signal has a shape of $330\times128\times2$, the SHAP gradient explainer, based on the expected gradients algorithm, computes corresponding Shapley values with a shape of $330\times128\times2\times11$, where 11 represents the number of modulation types in the network's output. The meaning of the Shapley value at position $(i,j,k,l)$ is:  The contribution of in-phase $(k=0)$ or quadrature $(k=1)$ part of its $j$th sampling data point with the $i$th sample classified into the $l$th modulation type.

For each sample, the Shapley values corresponding to the correct label are extracted, reducing the shape to $330\times128\times2$. Considering that the in-phase and quadrature parts of each data point are highly correlated and should not be considered independently, the sum of the Shapley values of these two components is used as the feature contribution for that data point. Summing over all samples while retaining the dimension of the data point dimension results in a one-dimensional array of length 128, where the $i$th Shapley value represents the overall contribution of the $i$th data point with respect to the dataset. A positive value indicates a positive contribution, while a negative value means that the feature contains negative information incurred by the attack, misleading the classification result. Finally, features are sorted based on their Shapley values, where smaller negative Shapley values correspond to features with a more significant deconstructive impact on the network (Instead of removing all destructive Shapley value features). Only the $m$ smallest negative Shapley values are selected as destructive feature points because some features may contribute destructively to one modulation type while being beneficial to another. Removing all destructive Shapley features would reduce the classification accuracy for previously correctly classified samples. Therefore, only the dominated destructive feature points are selected and removed.

The network is then fine-tuned using the refined attacked training and validation sets, from which the $m$ identified destructive feature points have been removed. The fine-tuning settings are as follows: the number of epochs is set to 60, the batch size is reduced to 20, the number of units in the first fully connected layer is decreased from 256 to 128, and early stopping is applied. Based on the indices of these destructive feature points, the corresponding features in the unattacked signals are also removed to eliminate residual negative information introduced by adversarial attacks. The processed signals are then fed into the fine-tuned AMC network for classification.

\subsection{Validation of SHAP}

\subsubsection{Consistency Between Shapley Values and Attack Levels}

\begin{figure}[htbp]
\centering
\includegraphics[width=0.45\textwidth]{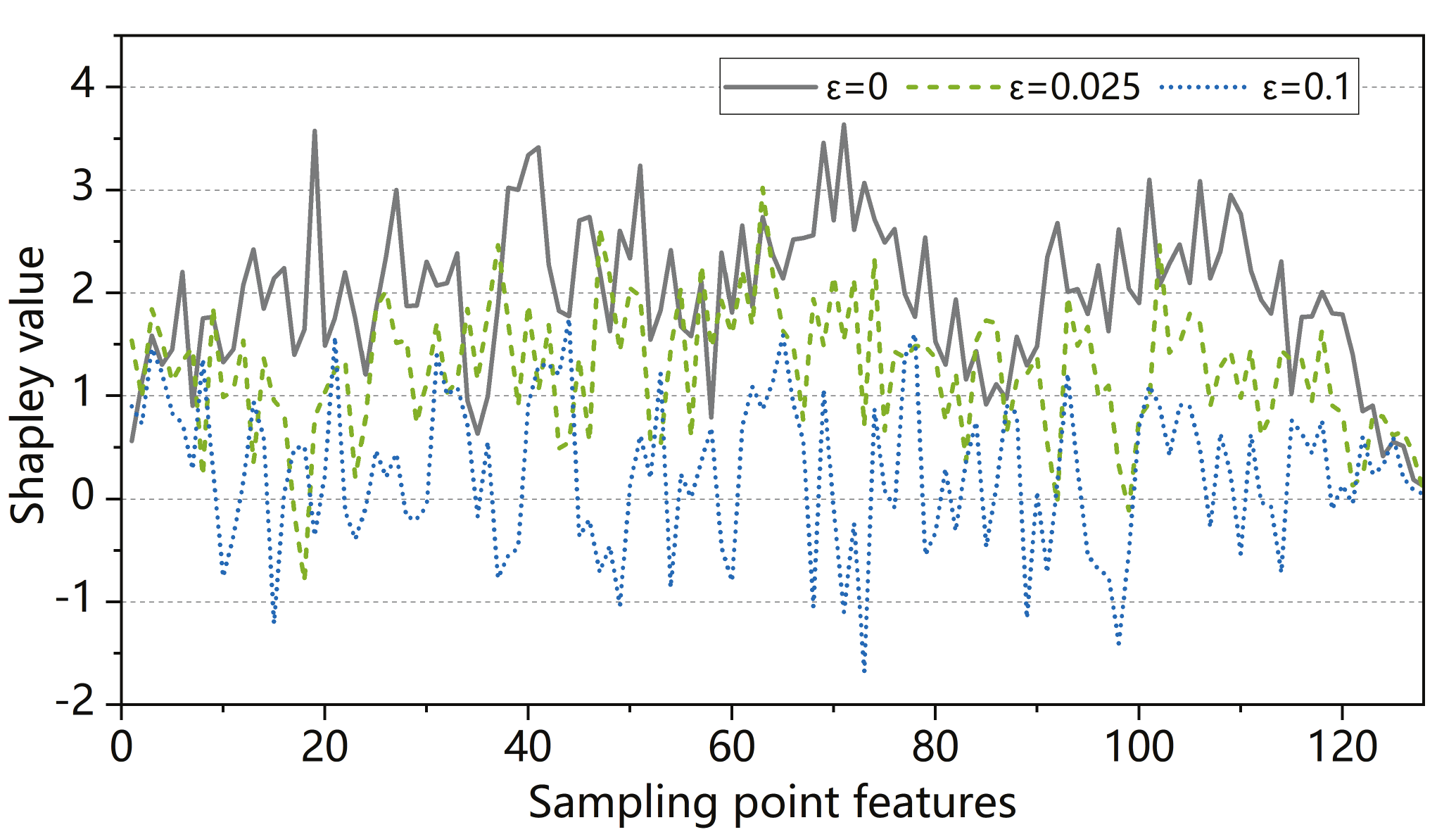}
\caption{Shapley values of the signal data points under different perturbation levels.}
\label{figneginfo}
\end{figure}

The PGD attack is applied to the labeled signals with perturbation levels $\varepsilon$ of 0, 0.025, and 0.1, respectively. Following the previously described data processing method, the Shapley summation values of 128 feature points are obtained, as illustrated in Fig. \ref{figneginfo}.
As the attack level increases, the Shapley values of the data points gradually decrease, with a growing number of values shifting from positive to negative. This trend clearly indicates that the higher attack level introduces more negative information into the input, leading to a degradation in modulation classification accuracy. These results demonstrate the effectiveness of the Shapley value as the key indicator.

\subsubsection{Consistency Between Shapley Value Heatmap and Classification Results}
\begin{figure}[!t]
\centering
\subfloat[Shapley heatmap.]{
	\includegraphics[scale=0.45]{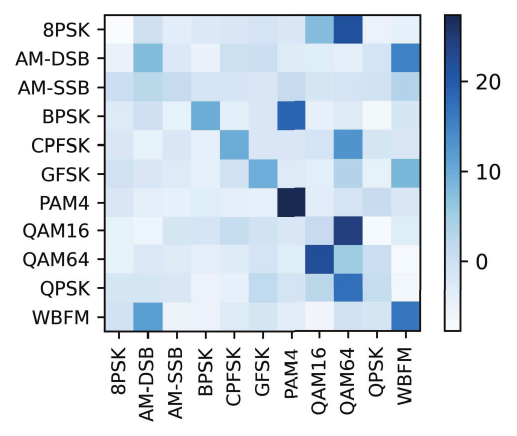}}
\subfloat[Confusion matrix.]{
	\includegraphics[scale=0.45]{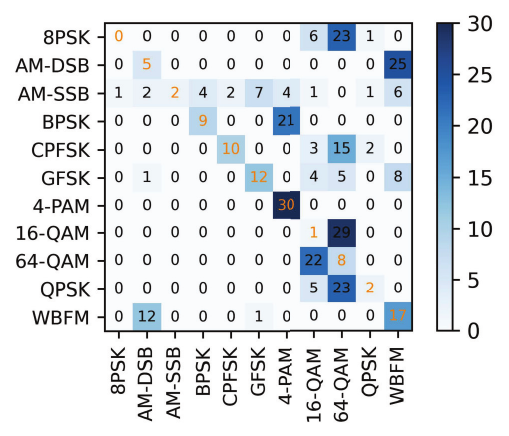}}
\caption{The heatmap of Shapley sum value and confusion matrix.}
\label{map_matx}
\end{figure}

The labeled signals are transformed into labeled attacked signals by applying a PGD attack with a perturbation level of 0.05. After computing the Shapley values for the labeled attacked signals, a heatmap is generated by aggregating the predicted Shapley values across all samples. This heatmap visualizes the distribution of Shapley values for each modulation type. In addition, a confusion matrix is plotted based on the classification results of the labeled attacked signals using the original (unmodified) AMC network, as shown in Fig. \ref{map_matx}.

The Shapley value at the $(i,j)$th position in Fig. \ref{map_matx}(a) is the sum of the Shapley values of all data points of all samples by predicting the $i$th modulation as the $j$th modulation. The count at the $(i,j)$th position of the confusion matrix in  Fig. \ref{map_matx}(b) is the number of samples predicting the $i$th modulation to be the $j$th modulation. The distribution consistency between two sub-figures reveal that the classification accuracy can be improved under the guidance of corresponding Shapley values.

\subsubsection{Classification Consistency Between Labeled and Unclassified Attacked Signals}
\begin{figure}[!t]
\centering
\subfloat[Labeled attacked signal.]{
	\includegraphics[scale=0.45]{Fig5b}}
\subfloat[Unclassified attacked signal.]{
	\includegraphics[scale=0.45]{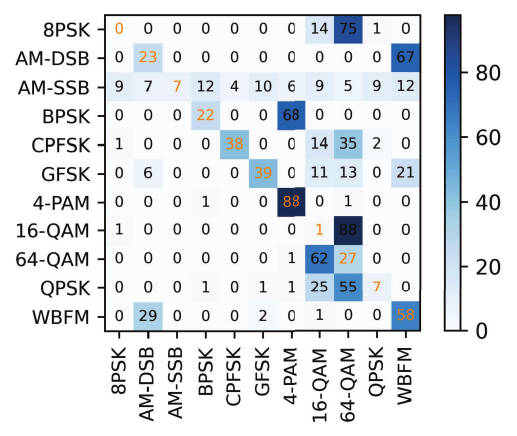}}
\caption{Comparison of confusion matrix.}
\label{figcomparison}
\end{figure}

Due to the label uncertainty of received attacked signals, direct SHAP analysis on realistic data is not practical. The confusion matrices after network prediction for labeled and unclassified attacked signals is shown in Fig. \ref{figcomparison}. It is shown that datasets with the similar distribution exhibit the similar classification result regardless of the size of datasets, which enables us to analyze the negative information as per the small dataset with known signal samples to guide the classification for the unknown attacked signals.

\subsection{Performance Evaluation and Analysis}

In this subsection, the classification performance of the proposed SHAP-AFT framework under FGSM, BIM, PGD, and C\&W attacks is evaluated.

\begin{figure}[t]
\centering
\includegraphics[width=0.42\textwidth]{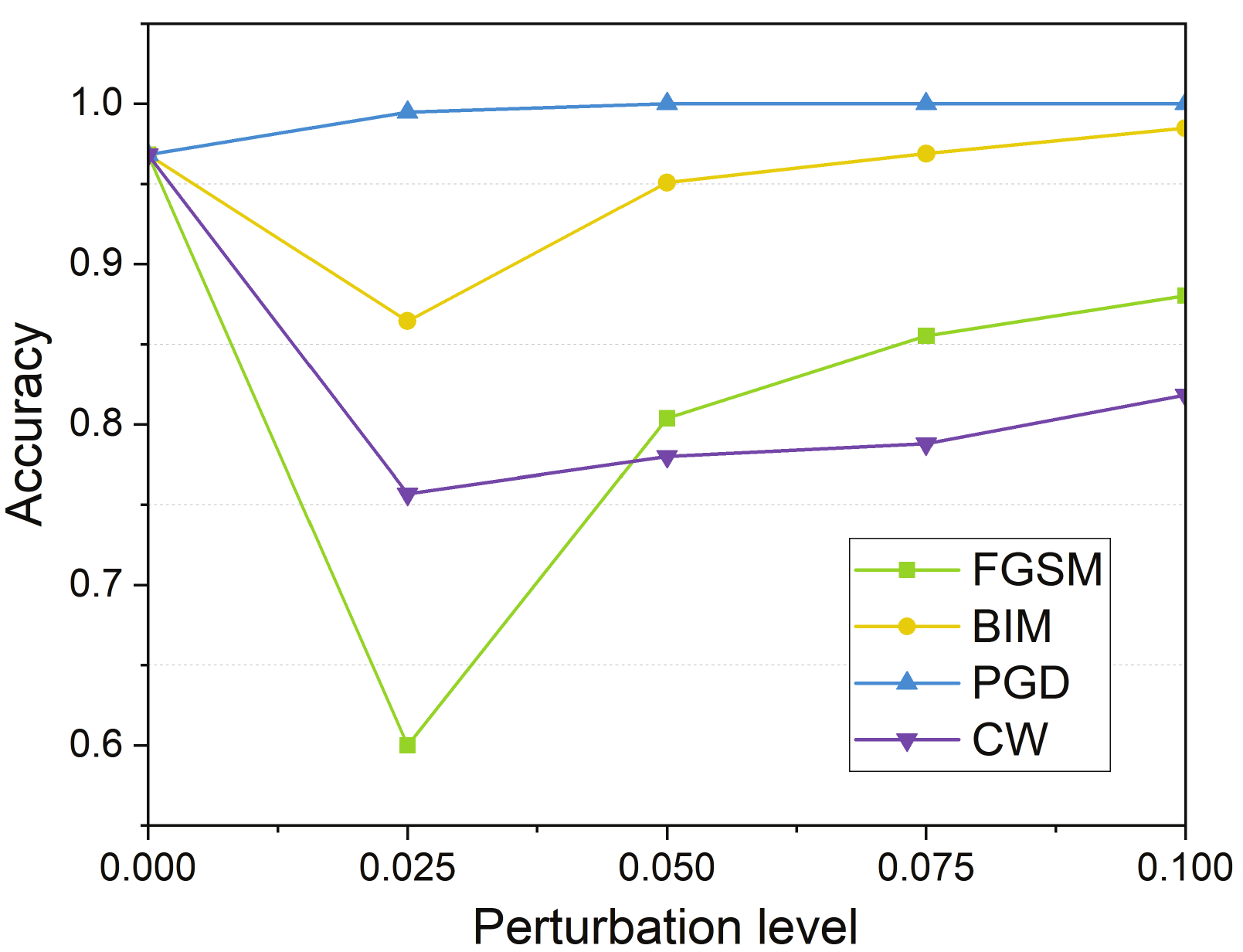}
\caption{The performance of attack detection network under different types of attack with different perturbation levels.}
\label{figdetectacc}
\end{figure}

The accuracy of the attack detection network on labeled signal samples under four types of attack and various perturbation levels is presented in Fig. \ref{figdetectacc}. A perturbation level of $\varepsilon=0$ corresponds to unattacked case. From Fig. \ref{figdetectacc}, the detection network achieves a high accuracy of about $97$\% in correctly identifying unattacked samples, allowing them to be directly forwarded to the original AMC network for classification without additional processing. The results also show that the detection network performs best for PGD attack with the detection accuracy approaching $100$\% at medium and high perturbation levels. In contrast, C\&W attack is detected with the lower accuracy since its attack mechanism is more sophisticated than gradient-based attacks, making the attack detection more difficult. Furthermore, the detection accuracy improves as the perturbation level increases, which aligns with the expectation that the higher attack level makes adversarial patterns more distinguishable.

\begin{figure}[t]
\centering
\includegraphics[width=0.45\textwidth]{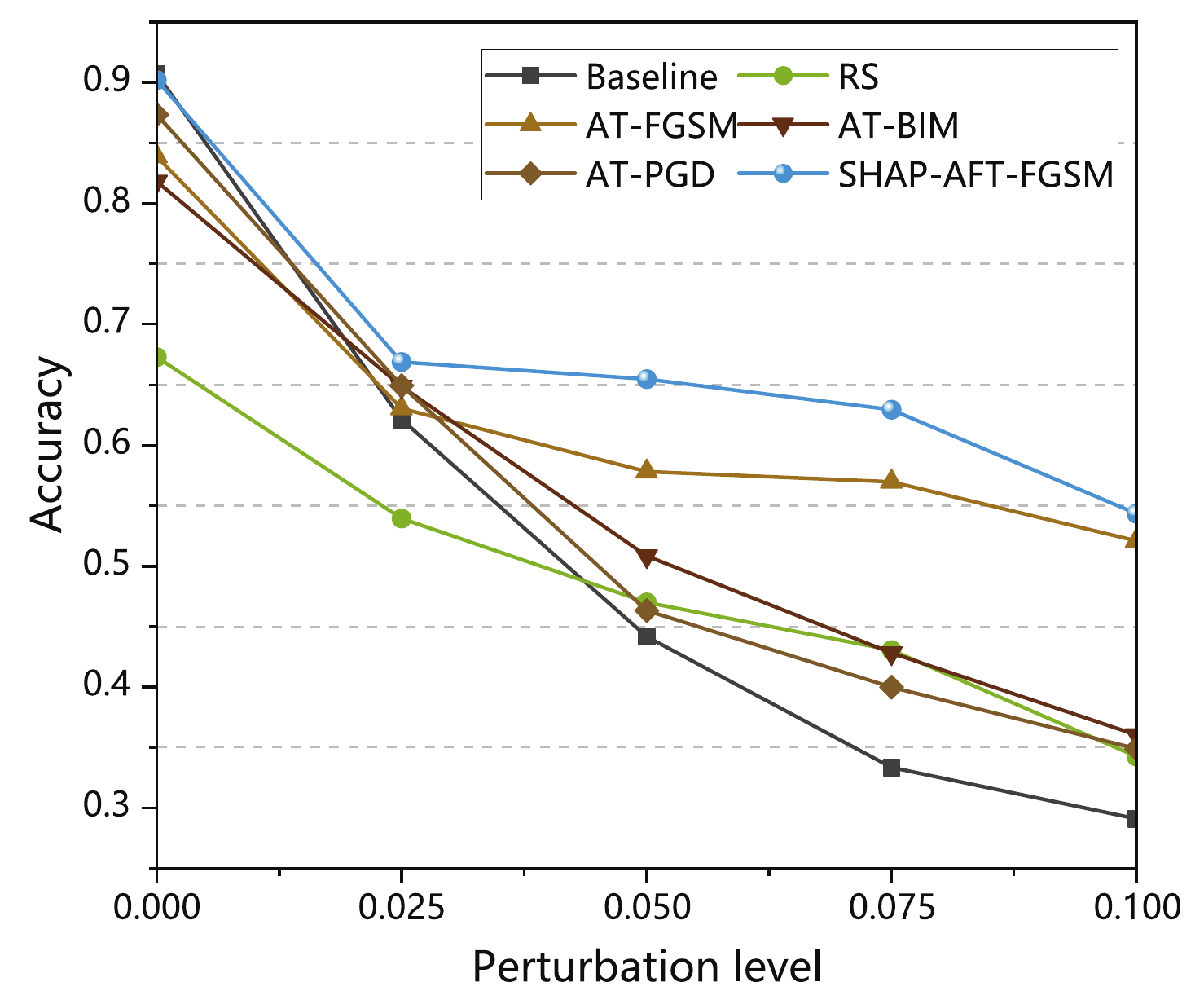}
\caption{The classification accuracy of SHAP-AFT under FGSM attack.}
\label{figfgsm}
\end{figure}
\begin{figure}[t]
\centering
\includegraphics[width=0.45\textwidth]{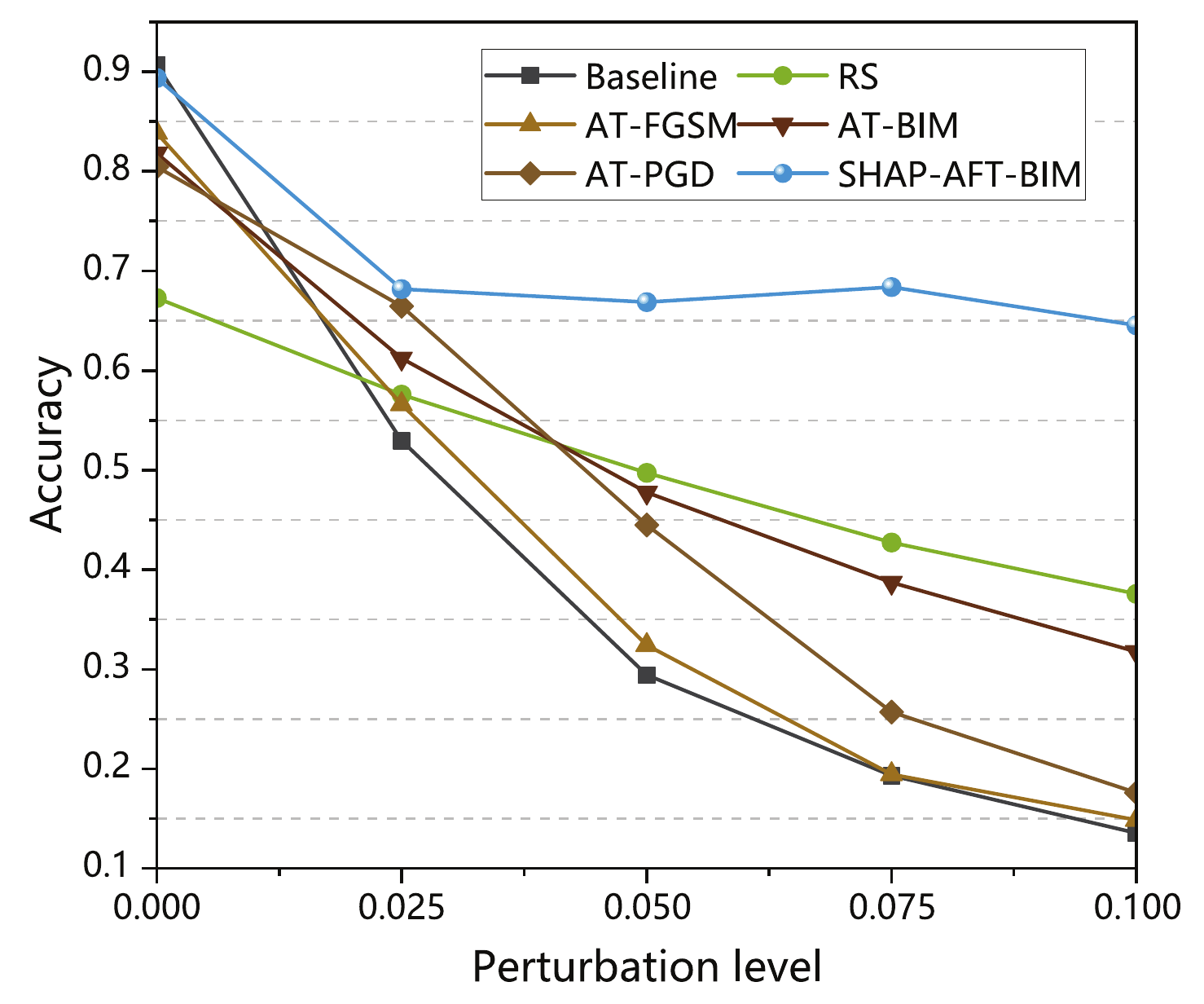}
\caption{The classification accuracy of SHAP-AFT under BIM attack.}
\label{figbim}
\end{figure}
\begin{figure}[t]
\centering
\includegraphics[width=0.45\textwidth]{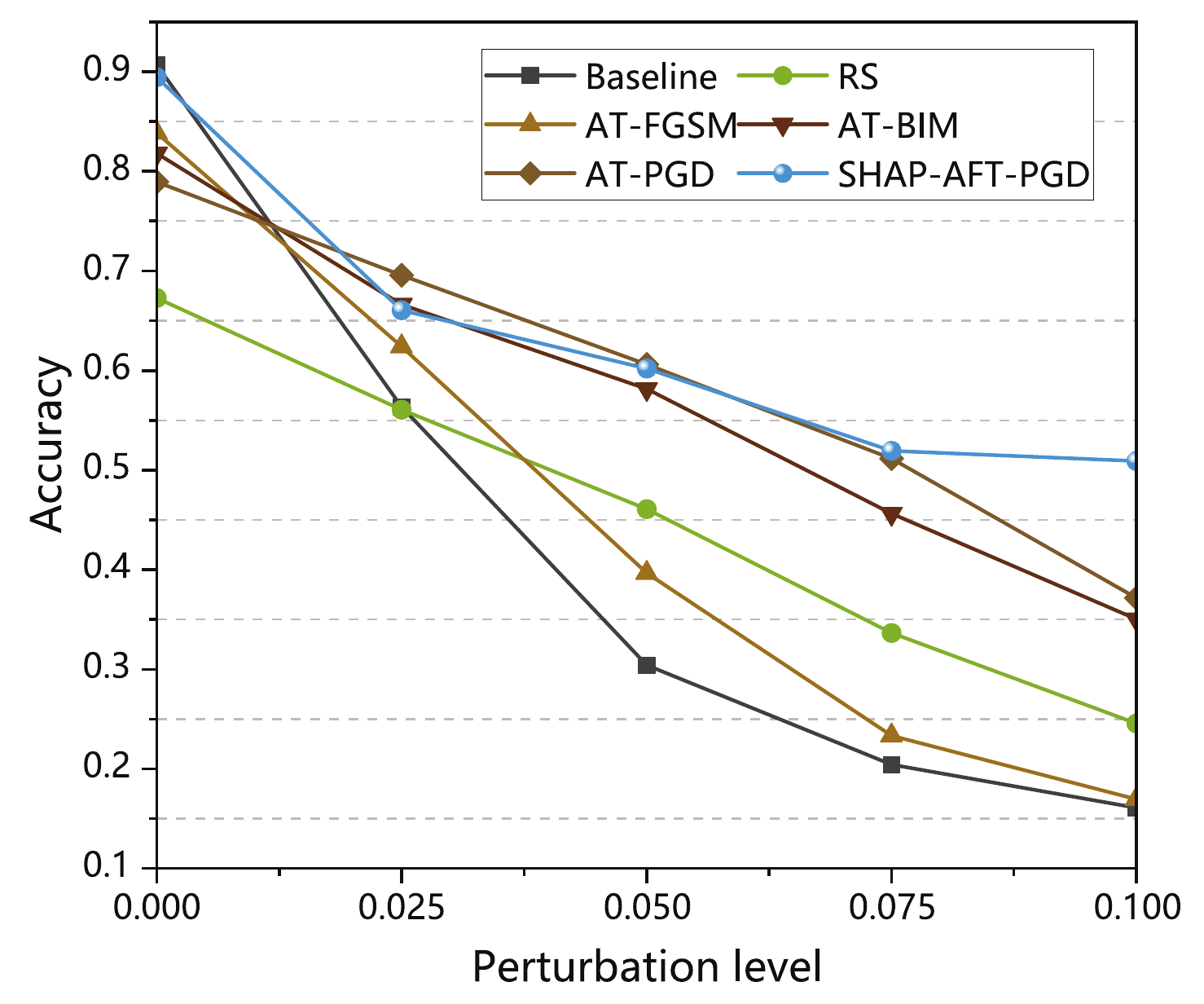}
\caption{The classification accuracy of SHAP-AFT under PGD attack.}
\label{figpgd}
\end{figure}

\begin{figure}[t]
\centering
\includegraphics[width=0.45\textwidth]{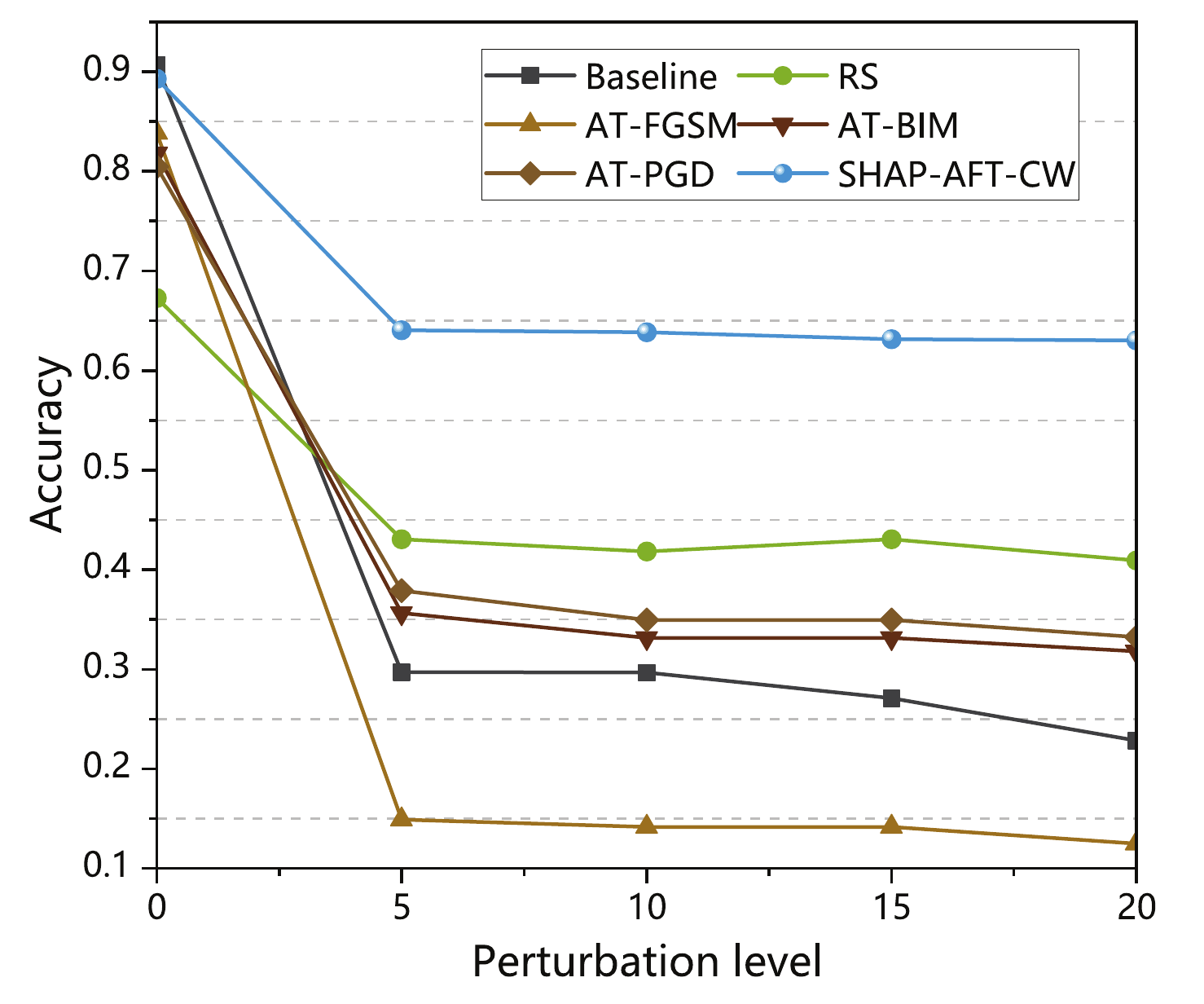}
\caption{The classification accuracy of SHAP-AFT under C\&W attack.}
\label{figcw}
\end{figure}

Figs.~\ref{figfgsm}--\ref{figcw} show the classification accuracy of the proposed SHAP-AFT under FGSM, BIM, PGD, and C\&W attacks, respectively. The existing defense schemes for comparison include adversarial training (AT) \cite{b8} using FGSM-, BIM-, PGD-attacked adversarial samples and random smoothing (RS) \cite{b11} with the added noise following a Gaussian distribution $\mathcal{N}(0, 0.02)$. 

For Figs.~\ref{figfgsm}--\ref{figpgd}, the classification performance is evaluated under four perturbation levels, i.e., $\varepsilon=0.025$, $0.05$, $0.075$, and $0.1$. For BIM attack, the perturbation step sizes corresponding to the four attack levels are set to $0.001$, $0.002$, $0.003$, and $0.004$, respectively. For PGD attack, the step size is fixed at 0.001. Both of them are conducted with $10$ iterations, and the random initialization is set as $0.3\varepsilon$ to make it no larger than $0.03$. For AT and SHAP-AFT, the adversarial samples are generated under FGSM, BIM, and PGD attacks with $\varepsilon=0.025$, yielding the corresponding schemes named AT-FGSM, AT-BIM, AT-PGD, SHAP-AFT-FGSM, SHAP-AFT-BIM, and SHAP-AFT-PGD, respectively. The ``Baseline" refers to the classification performance of the original AMC network. From these three figures, the proposed SHAP-AFT outperforms baseline schemes, especially with the medium and high $\varepsilon$. If the attack type used to generate adversarial samples is same as the actual attack type, AT can achieve the relatively good performance. From Fig.~\ref{figbim}, SHAP-AFT-BIM increases classification accuracy by about $15$\%, $38$\%, $49$\%, and $51$\% at the four perturbation levels, respectively, compared with the Baseline. From Fig.~\ref{figpgd}, although AT-PGD exhibits competitive performance with $\varepsilon \leq 0.025$, the proposed SHAP-AFT-PGD performs more stably as $\varepsilon$ increases to $0.1$. From Figs.~\ref{figfgsm}--\ref{figpgd}, compared with the Baseline, SHAP-AFT improves the accuracy by about $25$\%, $51$\%, and $34$\%, respectively, with $\varepsilon=0.1$, demonstrating its robustness to cope with the intense attack.

\begin{figure}[!t]
\centering
\subfloat[FGSM attack.]{
	\includegraphics[width=0.45\linewidth]{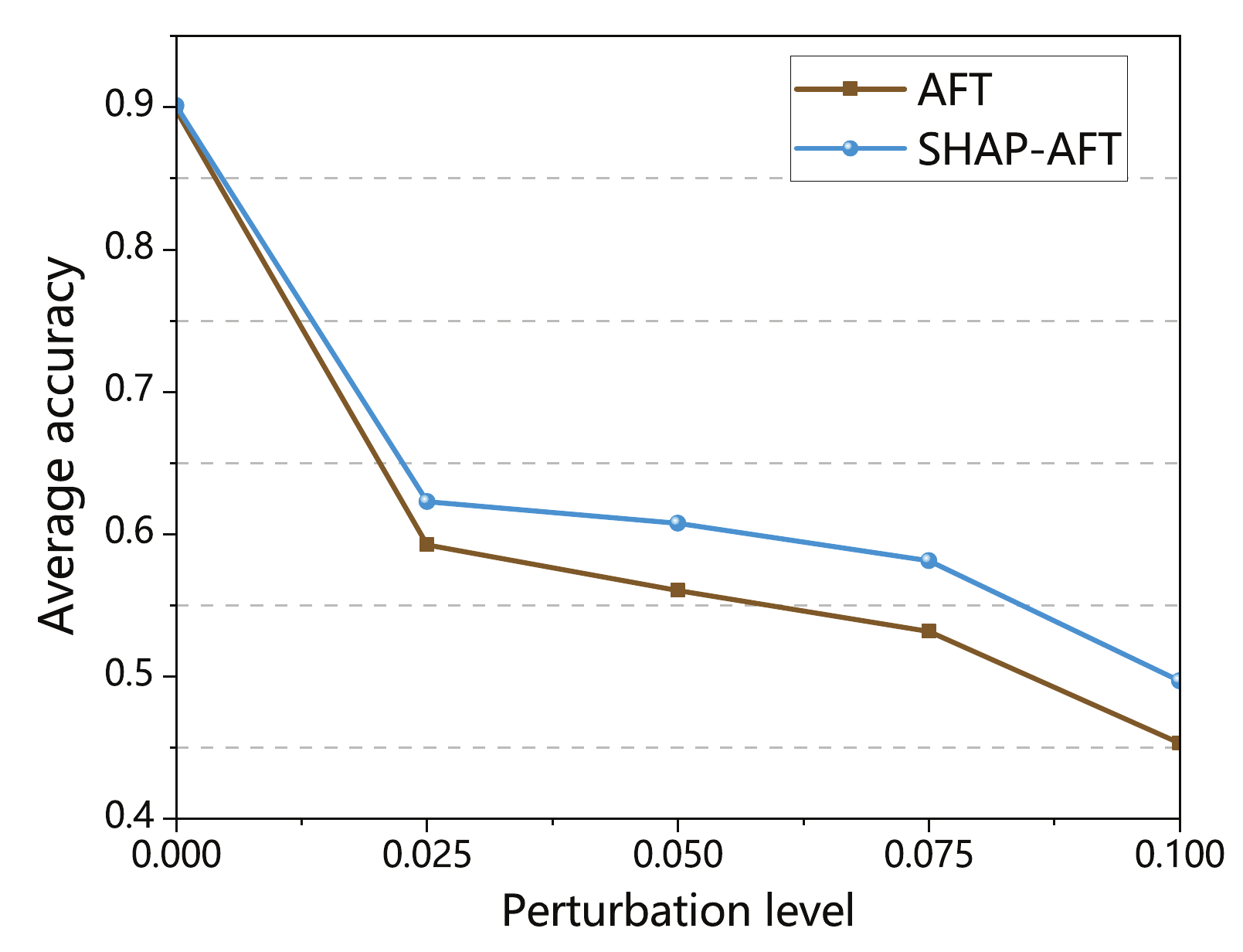}}
\subfloat[BIM attack.]{
	\includegraphics[width=0.45\linewidth]{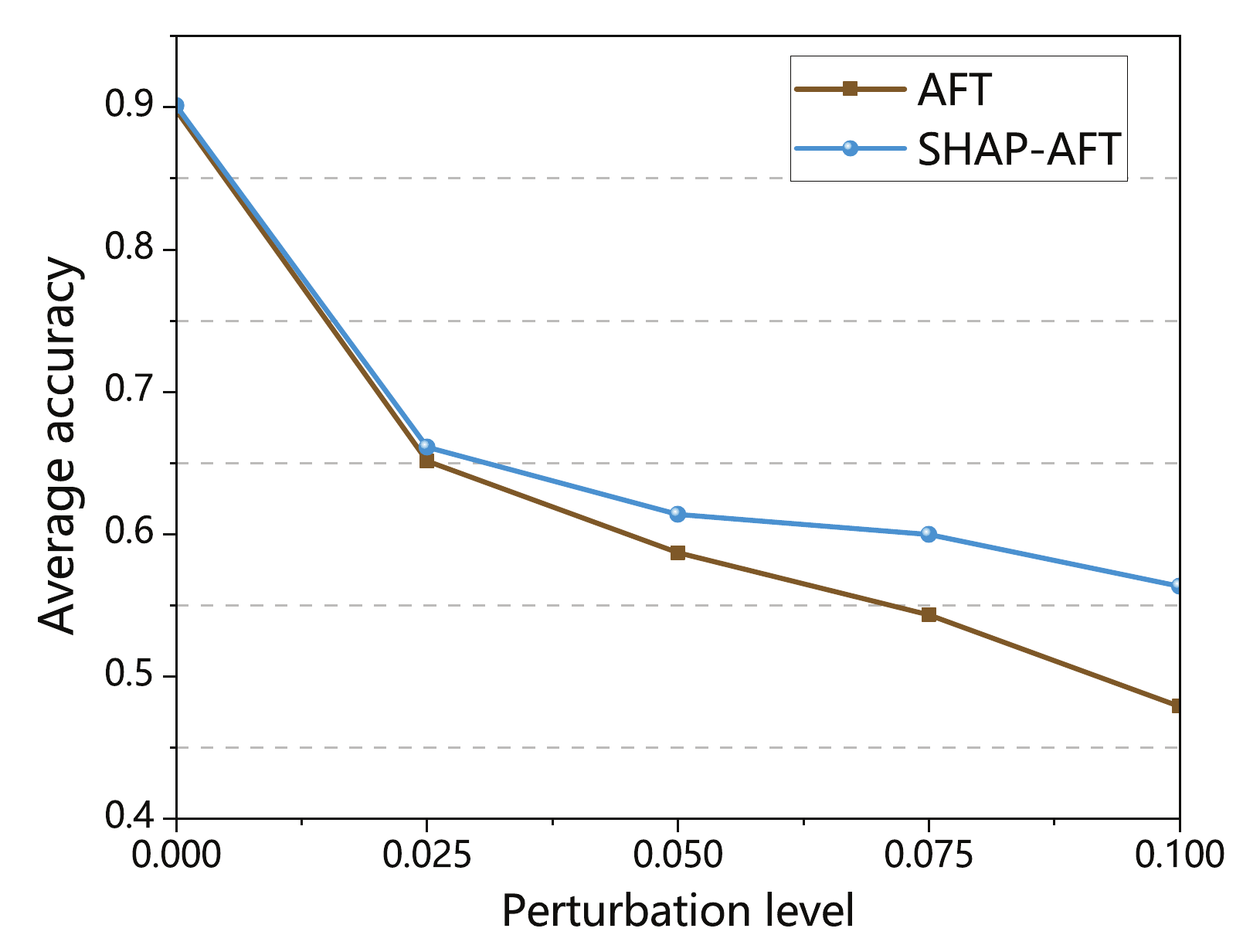}} \\
\subfloat[PGD attack.]{
	\includegraphics[width=0.45\linewidth]{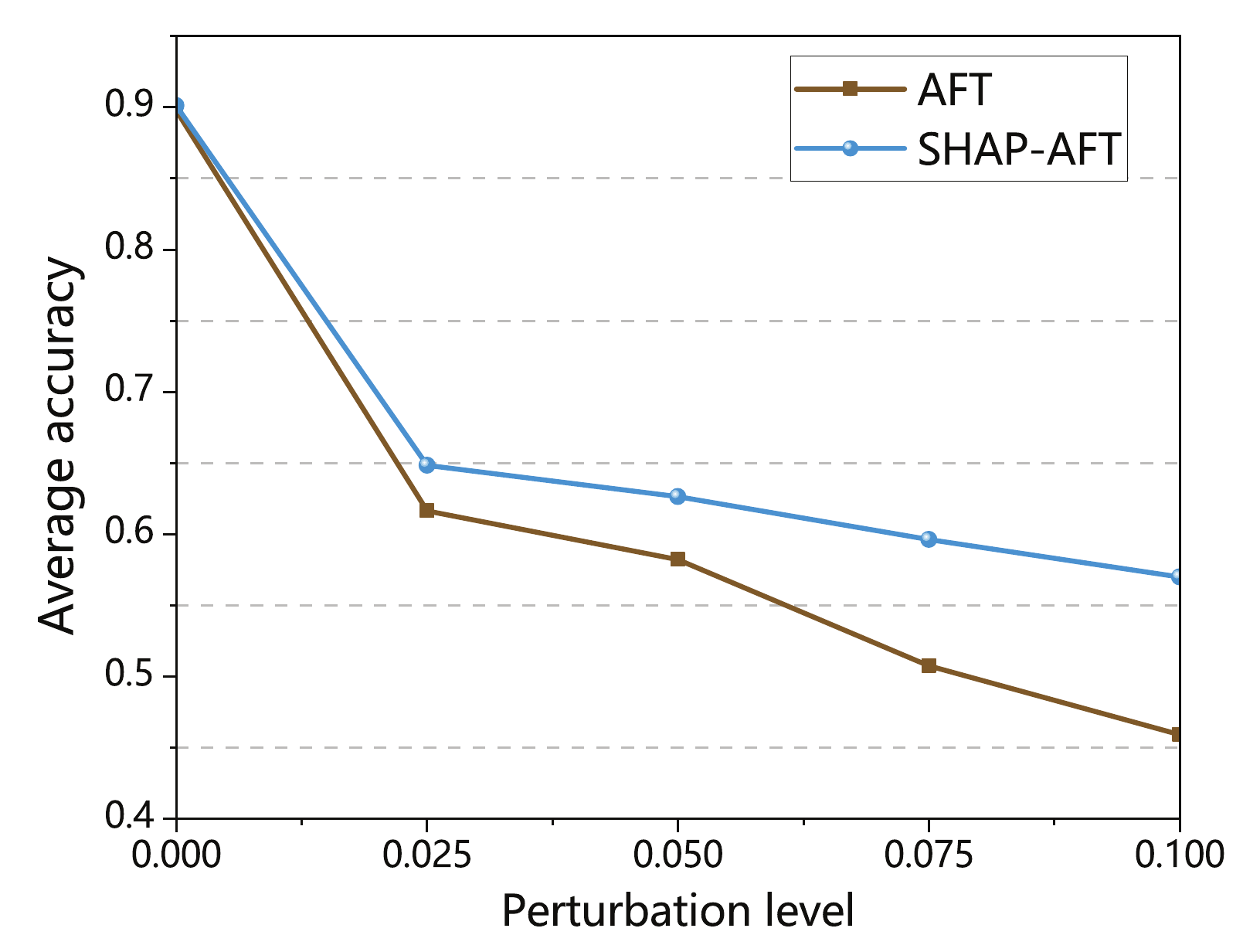}}
\subfloat[CW attack.]{
	\includegraphics[width=0.45\linewidth]{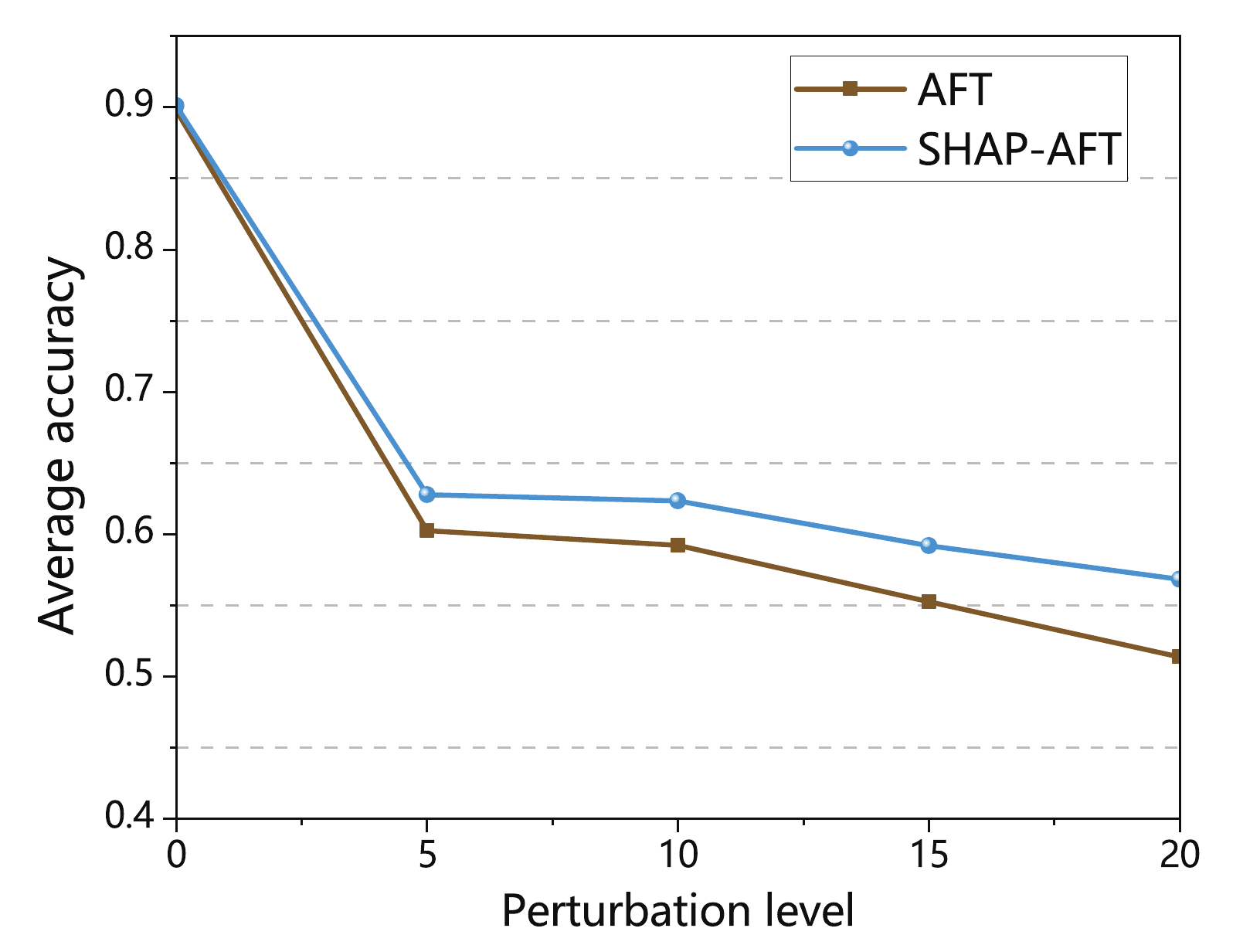}}
\caption{Average accuracy of AFT and SHAP-AFT with FT adversarial samples generated using four types of attacks.}
\label{figavg}
\end{figure}

For the optimization-based C\&W attack, the attacked background dataset, attacked training set, and attacked validation set used in the SHAP-AFT framework are generated under a C\&W attack with 5 iterations, resulting in the variant SHAP-AFT-CW. The C\&W attack is configured with an initial binary search value of 0.5, five binary search steps, and a confidence threshold of 0, representing its strongest attack setting. The optimizer’s learning rate is set to 0.03, and the performance is evaluated under four perturbation levels, i.e., maximum iteration counts of 5, 10, 15, and 20, respectively. The classification performance is presented in Fig. \ref{figcw}. Similar to those cases under gradient-based attacks in Figs.~\ref{figfgsm}--\ref{figpgd}, SHAP-AFT outperforms other baseline schemes by over $20$\% accuracy. However, RS stably achieves better performance than AT under the optimization-based C\&W attack. Therefore, SHAP-AFT is more versatile to cope with various types of attack compared with other defense schemes.

To further reveal the effect achieved by removing the negative information from adversarial samples, Fig.~\ref{figavg} shows the average accuracy of AFT and SHAP-AFT with FT adversarial samples generated under FGSM, BIM, PGD, and CW attacks, respectively. As an ablation of SHAP-AFT, AFT omits the negative information removing procedure. For each sub-figure, SHAP-AFT and AFT are tested under four types of attacks and the accuracy is averaged. From Fig.~\ref{figavg}, SHAP-AFT outperforms AFT significantly and the advantage becomes more pronounced as the attack intensity increases. For SHAP-AFT, using adversarial samples generated under PGD attack for FT achieves the best averaged accuracy.

\begin{figure}[htbp]
\centering
\includegraphics[width=0.45\textwidth]{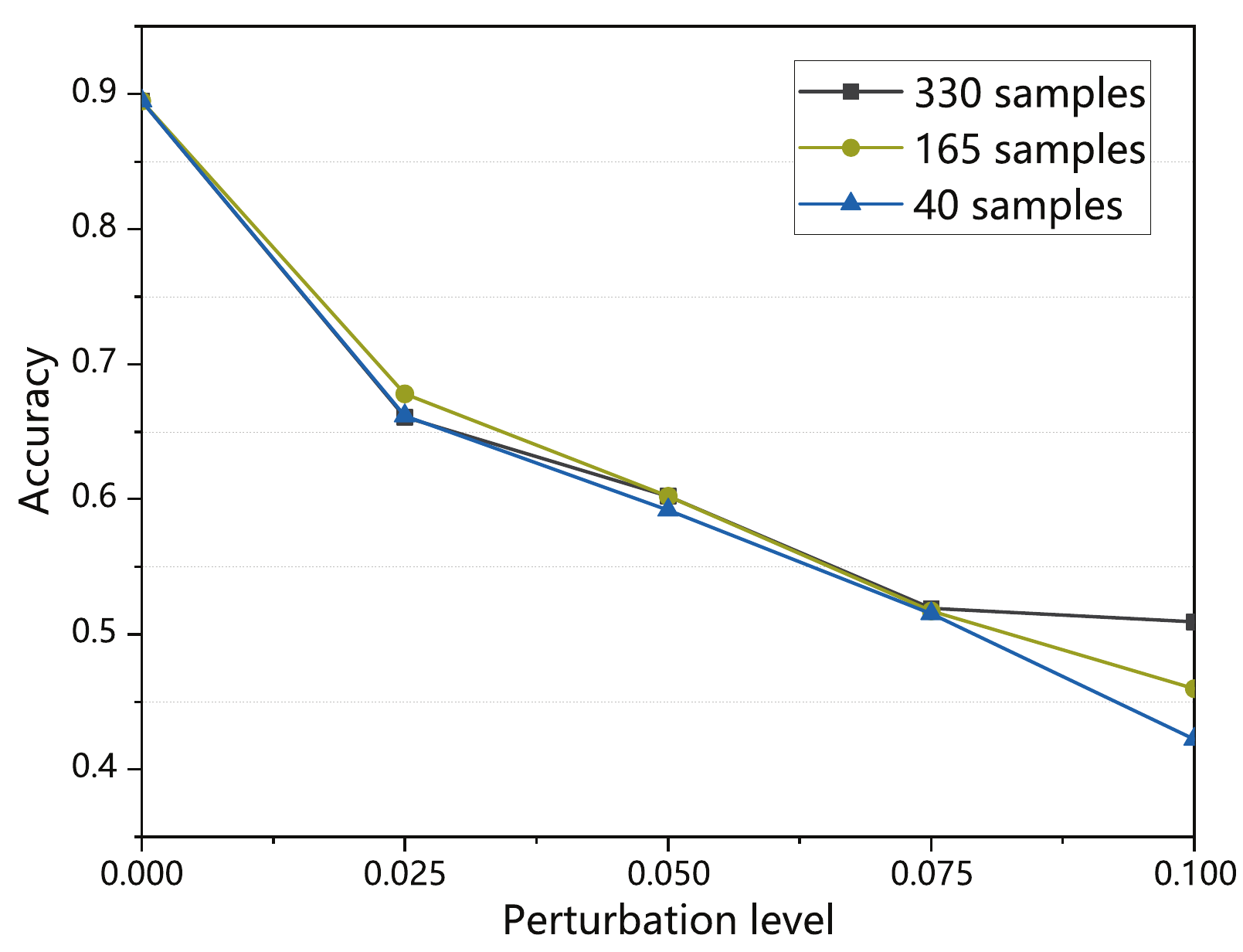}
\caption{Accuracy of SHAP-AFT with different numbers of labeled signal samples.}
\label{figtime}
\end{figure}

The labeled signal samples is utilized to compute Shapley values and the computation time is approximately proportional to the number of labeled samples. Using a larger number of labeled samples yields the better classification performance while increases the computation time. To investigate the tradeoff, Fig.~\ref{figtime} shows the accuracy of SHAP-AFT with $330$, $165$, and $40$ labeled signal samples, respectively. PGD attack is used in the adversarial sample generation and performance test. When the perturbation level $\varepsilon\leq 0.075$, different numbers of labeled samples yield almost the same accuracy. However, as $\varepsilon$ increases to $0.1$, the accuracies of $165$ and $40$ samples respectively decrease by about $4.95$\% and $8.69$\% compared with $330$ samples. Therefore, even if only limited number of labeled samples can be collected in practical applications, the classification performance will be not degraded significantly.

\section{Conclusion}
In this paper, an explainable deep learning-based adversarial defense framework, SHAP-AFT, is designed to improve the robustness and interpretability of AMC networks under adversarial attacks. A preliminary theoretical analysis is conducted to disclose the cognitive negative information embedded in attacked signals via the Shapley value decomposition. Based on this, the proposed SHAP-AFT realizes the adversarial defense using three stages. The first stage indicates the existence of the attack. The second stage evaluates contributions of the received data and removes those data positions with negative Shapley values corresponding to the dominating negative information caused by the attack. Finally, the AMC network is fine-tuned based on a small-scale adversarial dataset using the refined received data pattern. Simulation results verify the effectiveness of SHAP via three consistencies, followed by the overall classification performance demonstrating the superiority of the proposed SHAP-AFT.

\end{document}